\documentclass[aps,prl,twocolumn,showpacs,superscriptaddress,letterpaper,amsmath,amssymb]{revtex4}  

\usepackage{graphicx}
\usepackage{dcolumn}
\usepackage{bm}

\RequirePackage{xspace}
\usepackage{relsize}

\def\Kbar  {\kern 0.2em\overline{\kern -0.2em K}{}\xspace}

\def\Bbar    {\kern 0.18em\overline{\kern -0.18em B}{}\xspace}

\def\Qbar    {\kern 0.08em\overline{\kern -0.08em Q}{}\xspace}

\newcommand{\mev}{\ensuremath{\mathrm{\,Me\kern -0.1em V}}\xspace}
\newcommand{\mevc}{\ensuremath{{\mathrm{\,Me\kern -0.1em V\!/}c}}\xspace}
\newcommand{\mevcc}{\ensuremath{{\mathrm{\,Me\kern -0.1em V\!/}c^2}}\xspace}
\newcommand{\gev}{\ensuremath{\mathrm{\,Ge\kern -0.1em V}}\xspace}
\newcommand{\gevc}{\ensuremath{{\mathrm{\,Ge\kern -0.1em V\!/}c}}\xspace}
\newcommand{\gevcnospace}{\ensuremath{{\mathrm{\,Ge\kern -0.1em V\!/}c}}}
\newcommand{\gevcc}{\ensuremath{{\mathrm{\,Ge\kern -0.1em V\!/}c^2}}\xspace}

\begin{document}
\bibliographystyle{apsrev}

\title{{Search for High Mass Resonances Decaying to Muon Pairs in
    $\sqrt{s}=1.96$ TeV $p\bar{p}$ Collisions}}

\affiliation{Institute of Physics, Academia Sinica, Taipei, Taiwan 11529, Republic of China} 
\affiliation{Argonne National Laboratory, Argonne, Illinois 60439, USA} 
\affiliation{University of Athens, 157 71 Athens, Greece} 
\affiliation{Institut de Fisica d'Altes Energies, ICREA, Universitat Autonoma de Barcelona, E-08193, Bellaterra (Barcelona), Spain} 
\affiliation{Baylor University, Waco, Texas 76798, USA} 
\affiliation{Istituto Nazionale di Fisica Nucleare Bologna, $^z$University of Bologna, I-40127 Bologna, Italy} 
\affiliation{University of California, Davis, Davis, California 95616,
USA} 
\affiliation{University of California, Irvine, Irvine, California 92697, USA}
\affiliation{University of California, Los Angeles, Los Angeles, California 90024, USA} 
\affiliation{Instituto de Fisica de Cantabria, CSIC-University of Cantabria, 39005 Santander, Spain} 
\affiliation{Carnegie Mellon University, Pittsburgh, Pennsylvania 15213, USA} 
\affiliation{Enrico Fermi Institute, University of Chicago, Chicago, Illinois 60637, USA}
\affiliation{Comenius University, 842 48 Bratislava, Slovakia; Institute of Experimental Physics, 040 01 Kosice, Slovakia} 
\affiliation{Joint Institute for Nuclear Research, RU-141980 Dubna, Russia} 
\affiliation{Duke University, Durham, North Carolina 27708, USA} 
\affiliation{Fermi National Accelerator Laboratory, Batavia, Illinois 60510, USA} 
\affiliation{University of Florida, Gainesville, Florida 32611, USA} 
\affiliation{Laboratori Nazionali di Frascati, Istituto Nazionale di Fisica Nucleare, I-00044 Frascati, Italy} 
\affiliation{University of Geneva, CH-1211 Geneva 4, Switzerland} 
\affiliation{Glasgow University, Glasgow G12 8QQ, United Kingdom} 
\affiliation{Harvard University, Cambridge, Massachusetts 02138, USA} 
\affiliation{Division of High Energy Physics, Department of Physics, University of Helsinki and Helsinki Institute of Physics, FIN-00014, Helsinki, Finland} 
\affiliation{University of Illinois, Urbana, Illinois 61801, USA} 
\affiliation{The Johns Hopkins University, Baltimore, Maryland 21218, USA} 
\affiliation{Institut f\"{u}r Experimentelle Kernphysik, Karlsruhe Institute of Technology, D-76131 Karlsruhe, Germany} 
\affiliation{Center for High Energy Physics: Kyungpook National University, Daegu 702-701, Korea; Seoul National University, Seoul 151-742, Korea; Sungkyunkwan University, Suwon 440-746, Korea; Korea Institute of Science and Technology Information, Daejeon 305-806, Korea; Chonnam National University, Gwangju 500-757, Korea; Chonbuk National University, Jeonju 561-756, Korea} 
\affiliation{Ernest Orlando Lawrence Berkeley National Laboratory, Berkeley, California 94720, USA} 
\affiliation{University of Liverpool, Liverpool L69 7ZE, United Kingdom} 
\affiliation{University College London, London WC1E 6BT, United Kingdom} 
\affiliation{Centro de Investigaciones Energeticas Medioambientales y Tecnologicas, E-28040 Madrid, Spain} 
\affiliation{Massachusetts Institute of Technology, Cambridge, Massachusetts 02139, USA} 
\affiliation{Institute of Particle Physics: McGill University, Montr\'{e}al, Qu\'{e}bec, Canada H3A~2T8; Simon Fraser University, Burnaby, British Columbia, Canada V5A~1S6; University of Toronto, Toronto, Ontario, Canada M5S~1A7; and TRIUMF, Vancouver, British Columbia, Canada V6T~2A3} 
\affiliation{University of Michigan, Ann Arbor, Michigan 48109, USA} 
\affiliation{Michigan State University, East Lansing, Michigan 48824, USA}
\affiliation{Institution for Theoretical and Experimental Physics, ITEP, Moscow 117259, Russia}
\affiliation{University of New Mexico, Albuquerque, New Mexico 87131,
USA} 
\affiliation{New York University, New York, New York 10003, USA}
\affiliation{Northwestern University, Evanston, Illinois 60208, USA} 
\affiliation{The Ohio State University, Columbus, Ohio 43210, USA} 
\affiliation{Okayama University, Okayama 700-8530, Japan} 
\affiliation{Osaka City University, Osaka 588, Japan} 
\affiliation{University of Oxford, Oxford OX1 3RH, United Kingdom} 
\affiliation{Istituto Nazionale di Fisica Nucleare, Sezione di Padova-Trento, $^{aa}$University of Padova, I-35131 Padova, Italy} 
\affiliation{LPNHE, Universite Pierre et Marie Curie/IN2P3-CNRS, UMR7585, Paris, F-75252 France} 
\affiliation{University of Pennsylvania, Philadelphia, Pennsylvania 19104, USA}
\affiliation{Istituto Nazionale di Fisica Nucleare Pisa, $^{bb}$University of Pisa, $^{cc}$University of Siena and $^{dd}$Scuola Normale Superiore, I-56127 Pisa, Italy} 
\affiliation{University of Pittsburgh, Pittsburgh, Pennsylvania 15260, USA} 
\affiliation{Purdue University, West Lafayette, Indiana 47907, USA} 
\affiliation{University of Rochester, Rochester, New York 14627, USA} 
\affiliation{The Rockefeller University, New York, New York 10065, USA} 
\affiliation{Istituto Nazionale di Fisica Nucleare, Sezione di Roma 1, $^{ee}$Sapienza Universit\`{a} di Roma, I-00185 Roma, Italy} 

\affiliation{Rutgers University, Piscataway, New Jersey 08855, USA} 
\affiliation{Texas A\&M University, College Station, Texas 77843, USA} 
\affiliation{Istituto Nazionale di Fisica Nucleare Trieste/Udine, I-34100 Trieste, $^{ff}$University of Trieste/Udine, I-33100 Udine, Italy} 
\affiliation{University of Tsukuba, Tsukuba, Ibaraki 305, Japan} 
\affiliation{Tufts University, Medford, Massachusetts 02155, USA} 
\affiliation{University of Virginia, Charlottesville, VA  22906, USA}
\affiliation{Waseda University, Tokyo 169, Japan} 
\affiliation{Wayne State University, Detroit, Michigan 48201, USA} 
\affiliation{University of Wisconsin, Madison, Wisconsin 53706, USA} 
\affiliation{Yale University, New Haven, Connecticut 06520, USA} 
\author{T.~Aaltonen}
\affiliation{Division of High Energy Physics, Department of Physics, University of Helsinki and Helsinki Institute of Physics, FIN-00014, Helsinki, Finland}
\author{B.~\'{A}lvarez~Gonz\'{a}lez$^v$}
\affiliation{Instituto de Fisica de Cantabria, CSIC-University of Cantabria, 39005 Santander, Spain}
\author{S.~Amerio}
\affiliation{Istituto Nazionale di Fisica Nucleare, Sezione di Padova-Trento, $^{aa}$University of Padova, I-35131 Padova, Italy} 

\author{D.~Amidei}
\affiliation{University of Michigan, Ann Arbor, Michigan 48109, USA}
\author{A.~Anastassov}
\affiliation{Northwestern University, Evanston, Illinois 60208, USA}
\author{A.~Annovi}
\affiliation{Laboratori Nazionali di Frascati, Istituto Nazionale di Fisica Nucleare, I-00044 Frascati, Italy}
\author{J.~Antos}
\affiliation{Comenius University, 842 48 Bratislava, Slovakia; Institute of Experimental Physics, 040 01 Kosice, Slovakia}
\author{G.~Apollinari}
\affiliation{Fermi National Accelerator Laboratory, Batavia, Illinois 60510, USA}
\author{J.A.~Appel}
\affiliation{Fermi National Accelerator Laboratory, Batavia, Illinois 60510, USA}
\author{A.~Apresyan}
\affiliation{Purdue University, West Lafayette, Indiana 47907, USA}
% ...
\collaboration{CDF Collaboration\footnote{With visitors from $^a$University of Massachusetts Amherst, Amherst, Massachusetts 01003,
$^b$Istituto Nazionale di Fisica Nucleare, Sezione di Cagliari, 09042 Monserrato (Cagliari), Italy,
$^c$University of California Irvine, Irvine, CA  92697, 
$^d$University of California Santa Barbara, Santa Barbara, CA 93106
$^e$University of California Santa Cruz, Santa Cruz, CA  95064,
$^f$CERN,CH-1211 Geneva, Switzerland,
$^g$Cornell University, Ithaca, NY  14853, 
$^h$University of Cyprus, Nicosia CY-1678, Cyprus, 
$^i$University College Dublin, Dublin 4, Ireland,
$^j$University of Fukui, Fukui City, Fukui Prefecture, Japan 910-0017,
$^k$Universidad Iberoamericana, Mexico D.F., Mexico,
$^l$Iowa State University, Ames, IA  50011,
$^m$University of Iowa, Iowa City, IA  52242,
$^n$Kinki University, Higashi-Osaka City, Japan 577-8502,
$^o$Kansas State University, Manhattan, KS 66506,
$^p$University of Manchester, Manchester M13 9PL, England,
$^q$Queen Mary, University of London, London, E1 4NS, England,
$^r$Muons, Inc., Batavia, IL 60510,
$^s$Nagasaki Institute of Applied Science, Nagasaki, Japan, 
$^t$National Research Nuclear University, Moscow, Russia,
$^u$University of Notre Dame, Notre Dame, IN 46556,
$^v$Universidad de Oviedo, E-33007 Oviedo, Spain, 
$^w$Texas Tech University, Lubbock, TX  79609, 
$^x$Universidad Tecnica Federico Santa Maria, 110v Valparaiso, Chile,
$^y$Yarmouk University, Irbid 211-63, Jordan,
$^{gg}$On leave from J.~Stefan Institute, Ljubljana, Slovenia, 
}}
\noaffiliation

\begin{abstract}
 We present a search for a new narrow, spin-1, high mass resonance decaying to
  $\mu^+\mu^- + X $, using a matrix element based likelihood and a
  simultaneous measurement of the resonance mass and production
  rate. In data with 4.6 fb$^{-1}$ of integrated luminosity collected by the CDF detector in
  $p\bar{p}$ collisions at $\sqrt{s}=1960$ GeV, the most likely signal cross section is consistent with zero at 
  16\% confidence level.  We therefore do not observe evidence for 
  a high mass resonance, and place limits on models predicting 
  spin-1 resonances, including $M > 1071$ GeV/$c^2$ at 95\% confidence level for a 
  $Z'$ boson with the same couplings to fermions as the $Z$ boson.
\end{abstract}



\maketitle

\date{\today}

We report a search for a narrow spin-1 resonance ($Z'$) with decays to muon 
pairs and a mass between 130 GeV/$c^2$ and $\approx 1$ TeV/$c^2$.  Such a 
resonance is predicted generically in models with additional gauge groups 
\cite{langacker}, a feature of many extensions to the standard model of 
particle physics (SM).  Typical examples are Little Higgs models 
\cite{littlehiggs} and the next-to-minimal supersymmetric model \cite{nmssm}.  

Current 95\% confidence level (CL) lower limits on the mass of a $Z'$ boson with 
the same couplings to fermions as the $Z$ boson ($Z'_{SM}$) are $1030$ GeV/$c^2$ 
from a search of CDF dimuon data with 2.3 fb$^{-1}$ of integrated luminosity~\cite{zprimetemp}, and 
$1023$ GeV/$c^2$ \cite{d0ee} and $963$ GeV/$c^2$ \cite{cdfee} from respective 
searches in dielectron data from  D0 in 5.4 fb$^{-1}$ and from CDF in 2.5 fb$^{-1}$, respectively.
  
This Letter reports a new search of the CDF dimuon data with several significant 
enhancements:  twice the integrated luminosity, a matrix element based likelihood
providing an approximately 20\% relative increase in cross section
sensitivity at large $Z'$ mass, and a new 
statistical approach designed to maximize  simultaneously both discovery potential and mass 
exclusion limits in searches for new physics.  This new approach directly fits the data to a single
$Z'$ boson hypothesis in the plane of boson mass and signal cross
section, and can be applied to any search for a hypothetical new
particle of unknown mass.

We use a data sample corresponding to an integrated luminosity of 4.6 fb$^{-1}$, 
collected with the CDF II detector~\cite{cdf}, a general purpose detector designed 
to study $p\bar{p}$ collisions at the Fermilab Tevatron.  The tracking system 
consists of a cylindrical open-cell drift chamber and silicon microstrip detectors 
in a 1.4~T magnetic field parallel to the beam axis.  The silicon detectors provide 
tracking information for pseudorapidity $|\eta|<2$~\cite{coordinates} and are used 
to reconstruct collision and decay points.  The drift chamber surrounds the silicon 
detectors and gives full coverage in the central pseudorapidity region ($|\eta|<1$).  
For the muon kinematics relevant to this search, the tracker provides an invariant 
mass resolution of $\delta M / M^2 \approx 0.17$\%/TeV/$c^2$.  Electromagnetic and 
hadronic calorimeters surrounding the tracking system measure energy from particle 
showers and minimum ionizing particles such as muons.  Drift chambers and scintillators 
located outside the calorimeters detect muons in the central pseudorapidity region 
$|\eta|<1$.  

Events used for this search are selected online with a trigger requiring a muon 
candidate with $p_T > 18$ GeV/$c$.  The event selection is unchanged from the 
previous search~\cite{zprimetemp}, requiring at least two oppositely charged muons 
with $p_T > 30$ GeV$/c$ and no identified cosmic rays \cite{cosmics}.  Each muon 
is required to have calorimeter deposits consistent with a minimum ionizing particle 
and have a track fully within the fiducial volume of the drift chamber.  At least one muon must have 
an associated muon-chamber track.  

The dominant and irreducible background is standard model $Z/\gamma^*$ production
with subsequent decay to muons.  We model 
this background using events generated at leading order by {\sc pythia} \cite{pythia} with the 
{\sc cteq6l}~\cite{cteq} parton distribution functions (PDF).  Events are reweighted by the ratio of next-to-next-to-leading 
order to leading-order cross sections as a function of dimuon mass~\cite{carena}. 
We use the standard CDF simulation \cite{cdfsim} to model the detector response to 
the particles in the event.  The overall normalization of this background is derived 
from data near the $Z$ boson pole mass, $70 < M_{\mu\mu} < 110$ GeV$/c^2$.  

The remaining background contributions are small relative to standard
model sources of dimuons.  Decays of $WW$
(1\%) and $t\bar{t}$ (1\%) to 
dimuons are described using {\sc pythia} and the CDF detector simulation, normalized 
to NLO cross sections~\cite{ttandww}.  The background from objects
misidentified as muons (0.5\%) is 
estimated using the distribution of calorimeter energy in the vicinity
of each candidate muon in the sample~\cite{topdil}.  Cosmic ray
backgrounds ($<0.01\%$) are modeled using identified cosmic ray  events, weighted by the probability to survive the cosmic ray removal algorithm~\cite{cosmics}.

\begin{figure}[ht]
\begin{center}
\includegraphics[width=0.7\linewidth]{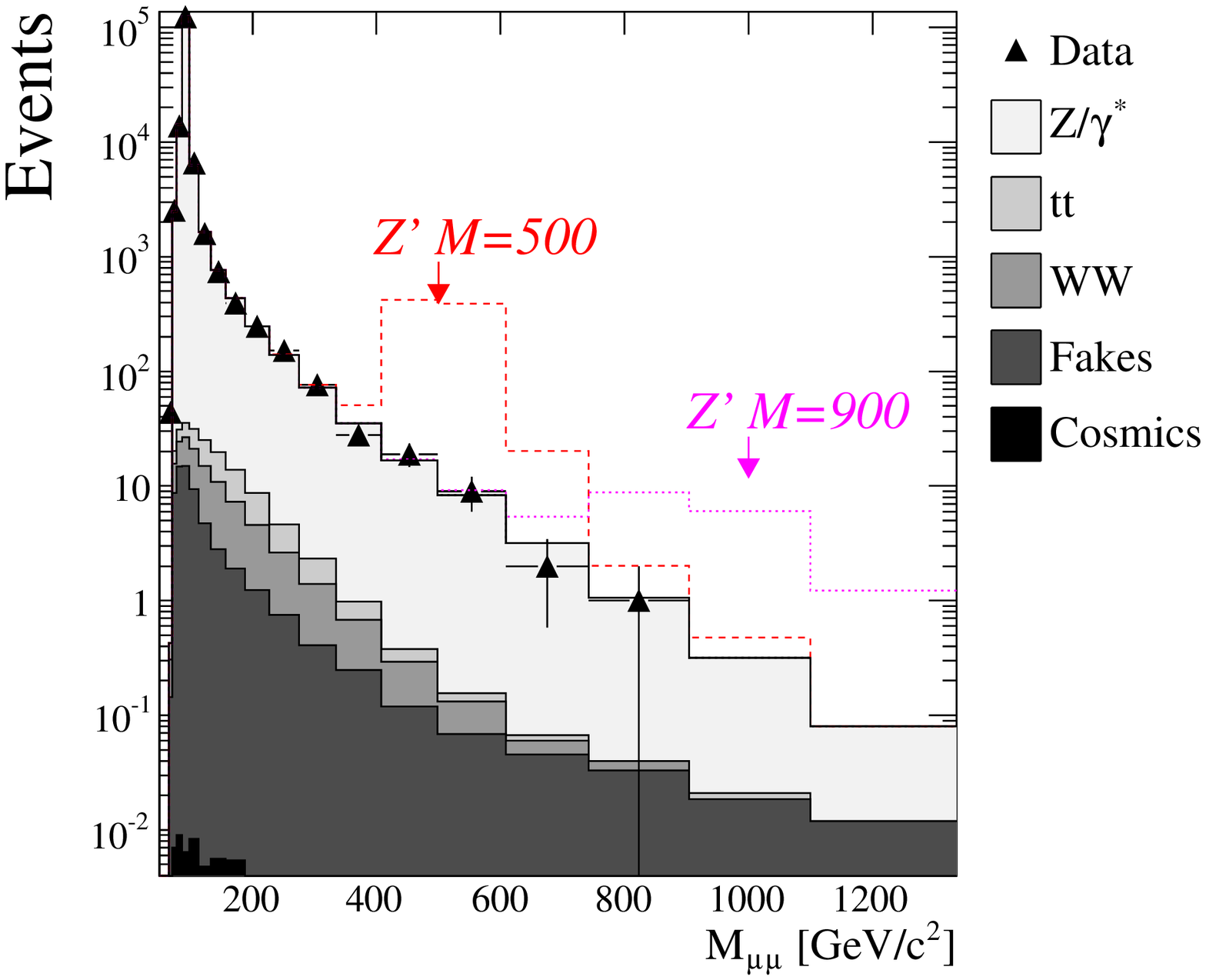}\\
\includegraphics[width=0.7\linewidth]{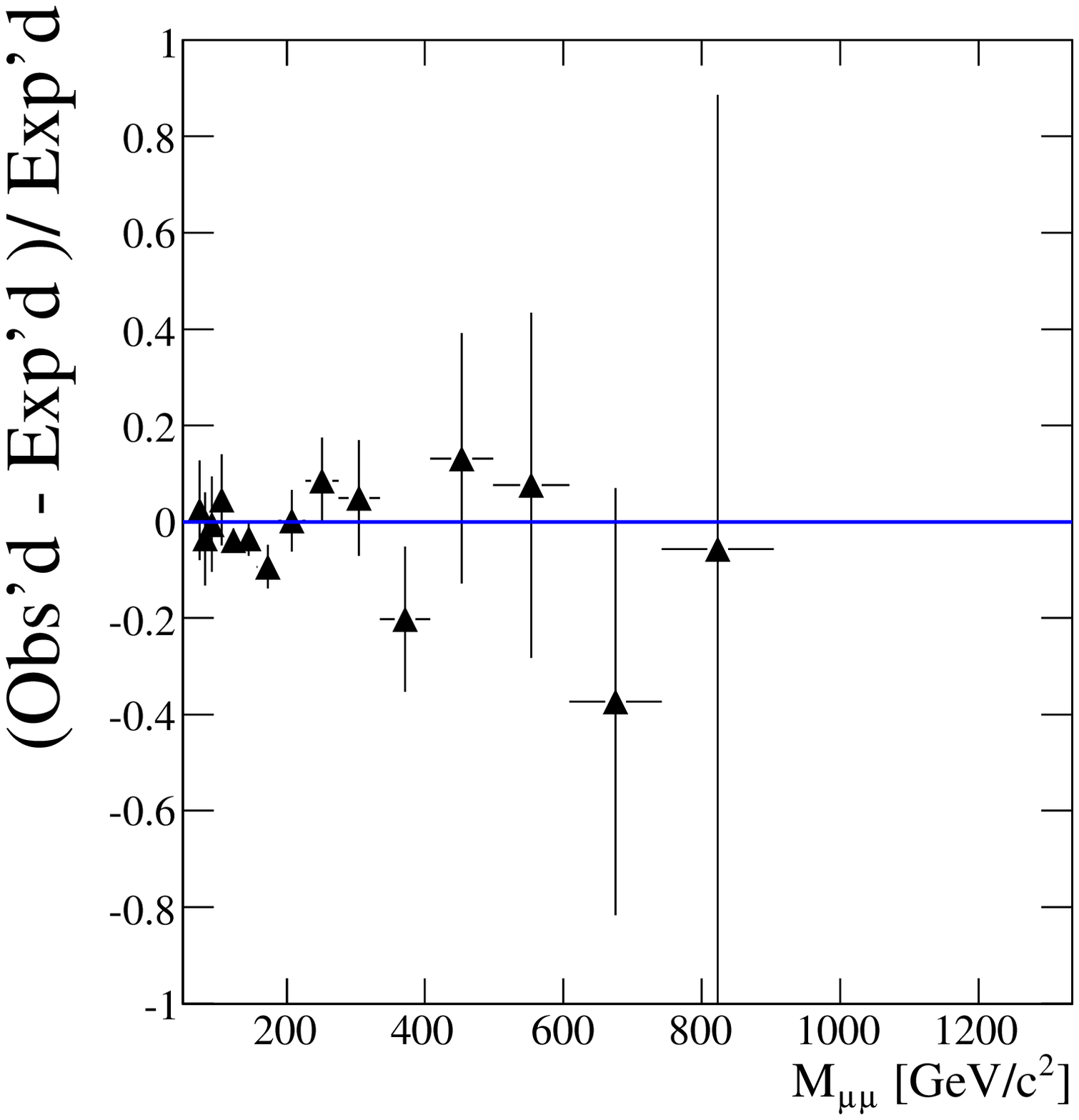}
\caption{ Top:  The distributions of $M_{\mu\mu}$ for data with 4.6
  fb$^{-1}$ of integrated luminosity
  (triangles) and expected SM backgrounds (histograms) with two
  example $Z'_{SM}$ signals.  Bottom: The relative 
  difference between observed and expected data, as a function of dimuon 
  mass. Error bars show statistical uncertainty.  }
\label{fig:mm}
\end{center}
\end{figure}

The observed spectrum of dimuon invariant mass ($M_{\mu\mu}$) is in
good agreement with the total SM plus cosmic-ray prediction, as shown
in Fig.~\ref{fig:mm}. We expect $1851 \pm 90$
events with $M_{\mu\mu} > 130$ GeV/$c^{2}$, and observe 1813 events.

We model the production and decay of a spin-1 resonance $Z'_{SM}$ using {\sc madevent}~\cite{madgraph}, assuming fermionic couplings equal to those of the $Z$ 
boson.  Initial-state QCD radiation, hadronization, and showering are modelled with 
{\sc pythia}.  The acceptance for $Z'$ boson events varies with mass, increasing 
from 20\% at 200 GeV$/c^2$ to nearly 40\% at 1 TeV/$c^2$, then dropping for 
$M_{Z'} > 1$ TeV/$c^2$ due to lack of initial-state partons with
sufficient momentum.

The previous CDF search \cite{zprimetemp} used a binned likelihood fit with bins uniform in the 
inverse of the reconstructed dimuon mass to extract the best-fit signal 
cross section at a range of masses.  That approach weighted events in the 
same reconstructed inverse mass bin equally, independent of expected track 
resolutions. We improve on this method by using an unbinned likelihood that includes
   a theoretical model of the full kinematics of the event and the
   event-by-event knowledge of the muon $p_T$ resolution. Well-measured events have narrower likelihoods, 
reflecting the higher quality of their information and making a stronger 
impact on the fitted result. The 
   likelihood is then used both to set limits on the $Z'$ cross section
   as a function of mass and to construct a 2D interval that gives a well-defined discovery
   condition.
 The likelihood is given by
\[ L( M_{Z'},s_{Z'}) = \frac{(N_{bg+Z'})^{N}
  e^{-(N_{bg+Z'})}}{N!} \prod^{N}_i L(x_i|M_{Z'},s_{Z'}), \]

\noindent
where $M_{Z'}$ is the $Z'$ pole mass, $N$ is the number
of events in the data, $N_{bg,Z'}$ are the number of events due to
background and $Z'$ signal, $x_i$ represents the observed kinematics of the $i$-th event, and $s_{Z'} = N_{Z'}/(N_{bg} + N_{Z'})$.  The dependence of the per-event likelihood is given by

\[ L(x_i|M_{Z'},s_{Z'}) =  s_{Z'}L_{Z'}(x_i|M_{Z'}) +  (1-s_{Z'})L_{Z/\gamma^*}(x_i), \]

\noindent
and the likelihood $L_{Z'}$ is calculated by integrating the
matrix element for $Z'$ production convolved with PDFs and the
detector resolution functions:
\[ L_{Z'}(x_i= p_1,p_2,\sigma_{p_{T1}},\sigma_{p_{T2}},N_{\rm
  jets}|M_{Z'}) = \]
\[\int d\Phi(q_1,q_2)
|\mathcal{M}_{Z'}(q_1,q_2,M_{Z'})|^2 f_{PDF}^{p} f_{PDF}^{\bar{p}} \]
\[ \times  \times T(p_1,q_1,\sigma_{p_{T1}})
T(p_2,q_2,\sigma_{p_{T2}}) \times P_{PT}(q_1+q_2,N_{\rm jets}), \]
\noindent
where $p_{1,2}$ represent the four-vectors of the two measured muons, $q_{1,2}$ represent the unknown four-vectors of the two true muons, $\Phi$ represents phase space for the true muons, 
$\mathcal{M}$ is the matrix element, $f_{PDF}$ is the
parton distribution function, $T(p,q,\sigma_{p_T})$ is the transfer function that parametrizes the detector
resolution as a function of the measured uncertainty $\sigma_{p_T}$, and $ P_{PT}$ is the probability density
function for $p_T$ of the $\mu\mu$ system, parameterized in the number
of jets ($N_{jets}$) with $E_T>15$ GeV and $|\eta|<2.5$.
The distributions of $\sigma_{p_T}$ for the $Z'$ signal and dominant
$Z/\gamma^*$ background are the same in the phase space region near the hypothesized $Z'$ mass, thus they do not affect the likelihood ratio ordering.  
 The distribution of $P_{PT}$ is obtained 
from simulated samples with initial- and final-state radiation.  An
analogous expression is used for $L_{Z/\gamma^*}$, which describes the
likelihood for the dominant $Z/\gamma^*$ background.

We analyze the resulting likelihood in two ways. First, we aim to
discover the regions in  $(M_{Z'},s_{Z'})$ that are consistent with the
  data and inconsistent with the SM, making no assumptions about the
  relationship between  $M_{Z'}$ and $s_{Z'}$. We refer to this as the
  {\it 2D interval} analysis. Second, we wish to set limits on the $Z'$
  mass in specific models.  In that case, we perform a  {\it
    raster scan}, in which we choose a set of values of $M_{Z'}$, and at each point derive limits on $s_{Z'}$. Together with a prediction for
  $s_{Z'}(M_{Z'})$ in a specific theory, we can use the raster scan to place
  lower limits on $M_{Z'}$.

\begin{figure}[ht]
\begin{center}
\includegraphics[width=0.7\linewidth]{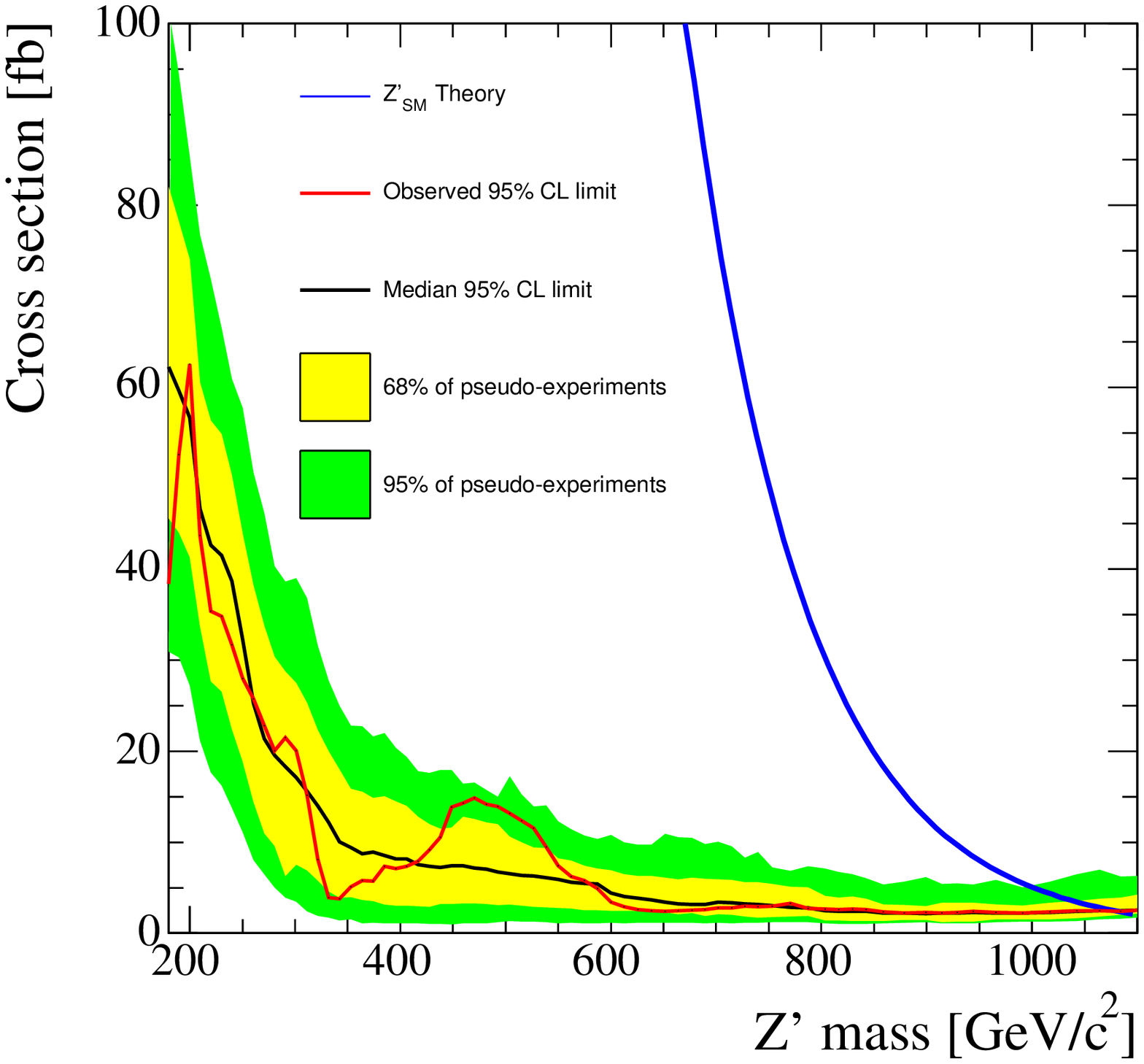}\\
\includegraphics[width=0.7\linewidth]{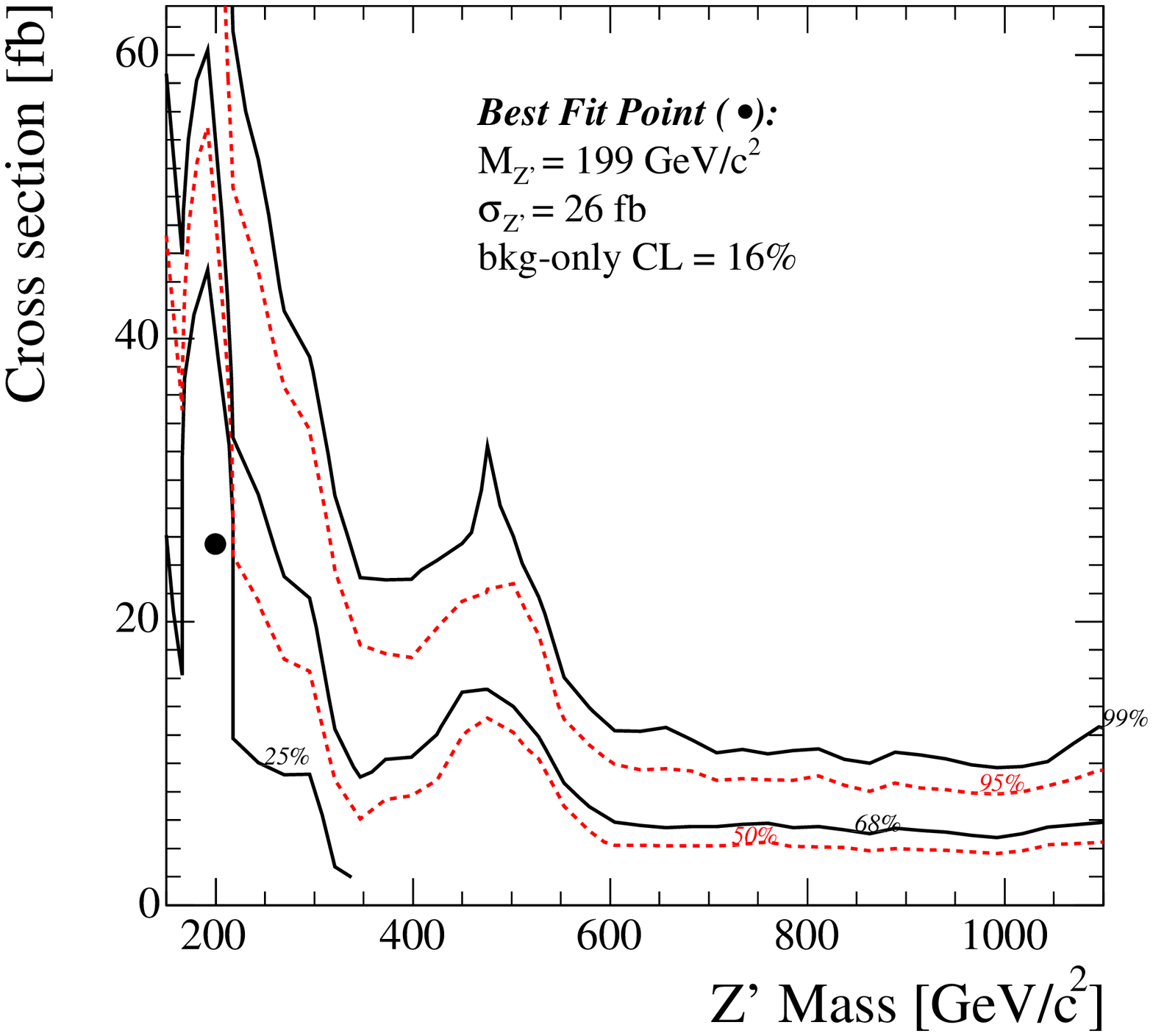}
\caption{ Observed 95\% CL limits in 1D raster scan (top) and 2D
  intervals (bottom)
  for data with 4.6 fb$^{-1}$ of integrated luminosity.  The solid circle indicates the best-fit value
  and the lines define regions in which 25, 50, 68, 95, and 99\%
  of experiments would yield results less consistent with
  the standard model and the data.}
\label{fig:4p6_obs}
\end{center}
\end{figure}

The 2D interval is constructed via the unified ordering
scheme~\cite{feldcous} in two
dimensions, resonance mass and cross section (see Fig.~\ref{fig:4p6_obs}).   At each test point in the
$(M_{Z'},s_{Z'})$ space, we calculate the ratio of the likelihood at the
test point to the likelihood at the best-fit point where the
likelihood is maximized.  To determine which test points are
consistent with the data at a given confidence level, we perform
pseudo-experiments to determine how often we expect to observe such a
likelihood ratio.  The pseudo-experiments include all backgrounds and
interference effects between the $Z$ and $Z'$, as well as variation of
the nuisance parameters from systematic uncertainties described below.   This approach is well
designed for discovery, as it tests the background
hypothesis exactly once. The significance of an observation corresponds
to the first confidence-level contour that includes a signal rate of
zero. It also provides a summary of $Z'$ models consistent with the
data without relying on specific model details. In the 2D interval analysis, the best-fit signal cross section of
$\sigma = 26$ fb occurs at  a resonance mass of $M = 199$ GeV/$c^2$,
but is consistent with zero at 16\% confidence level.

The raster scan is the traditional approach used in many analyses,
including the previous $Z'$ search~\cite{zprimetemp}.  It provides mass limits on theories 
that enforce a relationship between the signal fraction and the
resonance mass, and is appropriate if outside information 
indicates a particular mass is interesting.   In the presence of a significant excess
     above the background-only hyphothesis, one must account
     for the number of possible $Z'$ masses, each of which tests the
     background-only hypothesis: the {\it look-elsewhere effect}. The
     2D analysis needs no such correction.  While the resulting
     discovery significance of the corrected raster scan will be correct, 
  one will still be left with an interval in the signal fraction at
  every mass.  In contrast, in the presence of a signal, the
     2D analysis will provide a range of masses that is consistent
     with the signal.

%

\begin{figure}[t]
\begin{center}
\includegraphics[width=0.95\linewidth]{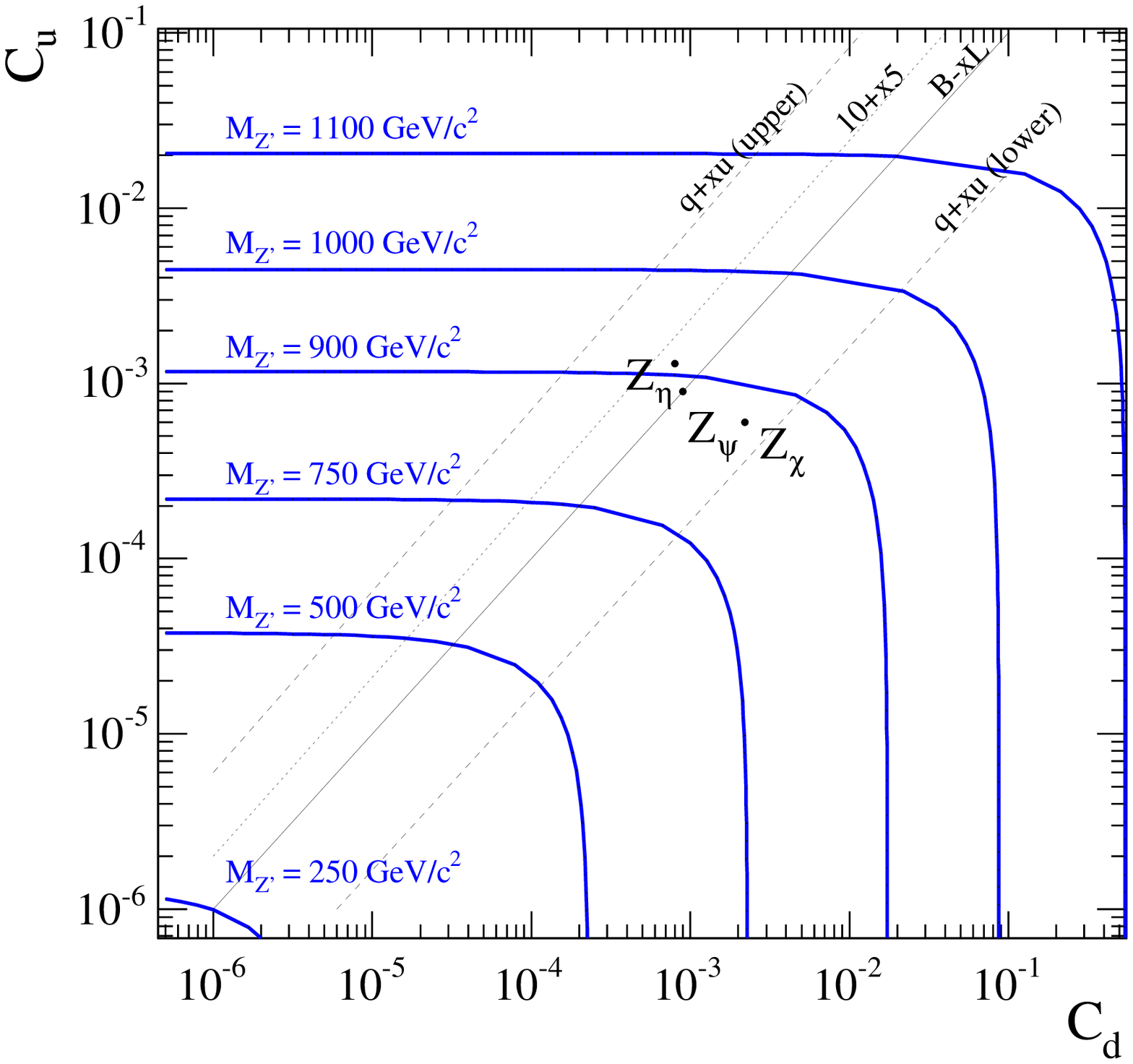}
\caption{ Observed 95\% CL limits for data with 4.6 fb$^{-1}$ of
  integrated luminosity expressed
  as limits on the up and down type charges $c_u$ and
  $c_d$~\cite{carena}. The solid and dotted lines show possible models
in $U(1)_{B-XL}$ and $U(1)_{10+x\bar{5}}$ groups, respectively. The
dashed lines show the range for models in the $U(1)_{q+xu}$ group.}
\label{fig:4p6_obs_lim}
\end{center}
\end{figure}

%

\begin{table}[h]
\begin{center}
\caption{Mass limits on specific spin-1 $Z'$ models~\cite{carena} in
  data with 4.6 fb$^{-1}$ of integrated luminosity at 95\% confidence level.}
\begin{ruledtabular}
\begin{tabular}{lrrrrrrr}
Model & $Z'_l$& $Z'_{sec}$& $Z'_N$& $Z'_\psi$& $Z'_\chi$& $Z'_\eta$& $Z'_{SM}$\\
Mass Limit (GeV/$c^2$) & 817 & 858 & 900 & 917 & 930 & 938 &  1071\\
\end{tabular}
\end{ruledtabular}

\label{tab:limits}
\end{center}
\end{table}

Dominant systematic uncertainties~\cite{zprimetemp} include
uncertainties on the PDFs and the
dependence of the next-to-leading order cross section on the dimuon
invariant mass. These weaken the final limits by 5-10\% depending on
mass. Additional uncertainties are the level of initial state
radiation and muon acceptance at large transverse momentum.

The raster scan in mass allows us to set strong limits on specific
models of $Z'$ production; see Fig.~\ref{fig:4p6_obs} and Table
~\ref{tab:limits}. The production cross section times
branching fraction to the dimuon final state is determined by the
couplings of the fermions     to the
$Z'$. Figure~\ref{fig:4p6_obs_lim} shows how mass limits depend on the
charges     of the up- and down-type fermions to the U(1) group
associated      with the $Z'$. Table ~\ref{tab:limits} shows the
limits for the specific models described  in Ref~\cite{carena}.

In conclusion, we have applied the matrix-element-based likelihood
technique to a search for new spin-1 resonances decaying to muon
pairs, set the strongest limits to date on the resonance
cross section and mass, and introduced a statistical analysis approach
that is useful for this
analysis as well as for potential LHC discoveries.

\begin{acknowledgments} We acknowledge useful conversations with Tim Tait.
We thank the Fermilab staff and the technical staffs of the participating institutions for their vital contributions. This work was supported by the U.S. Department of Energy and National Science Foundation; the Italian Istituto Nazionale di Fisica Nucleare; the Ministry of Education, Culture, Sports, Science and Technology of Japan; the Natural Sciences and Engineering Research Council of Canada; the National Science Council of the Republic of China; the Swiss National Science Foundation; the A.P. Sloan Foundation; the Bundesministerium f\"ur Bildung und Forschung, Germany; the World Class University Program, the National Research Foundation of Korea; the Science and Technology Facilities Council and the Royal Society, UK; the Institut National de Physique Nucleaire et Physique des Particules/CNRS; the Russian Foundation for Basic Research; the Ministerio de Ciencia e Innovaci\'{o}n, and Programa Consolider-Ingenio 2010, Spain; the Slovak R\&D Agency; and the Academy of Finland.
\end{acknowledgments}


\begin{thebibliography}{99}
\bibitem{coordinates}
CDF uses a cylindrical coordinate system with
  the $z$ axis along the proton beam axis. Pseudorapidity is $\eta
  \equiv -\ln(\tan (\theta/2))$, where $\theta$ is the polar angle
  relative to the proton beam direction, and $\phi$ is the azimuthal
  angle  while $p_T = |p|\ \sin\theta$, $E_T = E\ \sin\theta$.
\end{thebibliography}
\end{document}